\documentstyle[12pt,aasms]{article}
\tightenlines
\begin{document}
\def\ltsima{$\; \buildrel < \over \sim \;$}\
\def\simlt{\lower.5ex\hbox{\ltsima}}
\def\gtsima{$\; \buildrel > \over \sim \;$}
\def\simgt{\lower.5ex\hbox{\gtsima}}
\def\etal{{\it et al.~}}
\def\minspt{$^{\prime}_\cdot$}
\def\secspt{$^{\prime\prime}_\cdot$}
 
\title { Emission-Line Properties of 3CR Radio Galaxies III: \\
          Origins and Implications of the Velocity Fields }

\author{ Stefi A. Baum\altaffilmark{1} \& Patrick J. McCarthy\altaffilmark{2,3}}

\altaffiltext{1}{The Space Telescope Science Institute, \\
                 3700 San Martin Dr., Baltimore, MD 21218 }

\altaffiltext{2}{The Observatories of the Carnegie Institution of Washington, \\
                  813 Santa Barbara St., Pasadena, CA 91101}
 
\altaffiltext{3}{Guest Observer at the National Optical Astronomy
Observatories, Cerro Tololo Interamerican Observatory,
which is operated by the Associated Universities for Research in
Astronomy, Inc., under contract with the National Science Foundation}

\begin{abstract}

   We present the results of an analysis of the large scale velocity
fields of the ionized gas associated with powerful radio galaxies.
Long-slit spectra of 52 objects provide a sample of resolved velocities
that span a wide range of redshifts, radio and emission-line luminosities.
Line widths reaching 1000 km s$^{-1}$ and resolved velocity fields with
amplitudes of up 1500 km s$^{-1}$ are found on scales from 10 to 100
kpc in the environments of radio galaxies at redshifts larger than 0.5.
The global velocities and FWHM are of comparable amplitudes in the FRII
sources, while the FRI sources have FWHM values that are larger than
their resolved velocity fields.  We find evidence for systematically
larger line widths and velocity field amplitudes at $z > 0.6$.  Several
of the largest amplitude systems contain two galaxies with small
projected separations. All of the $> 1000$ km s$^{-1}$ systems occur in
objects at $z > 0.6$ and all have comparable radio and [OII] sizes.  
There is a weak correlation of off-nuclear line widths 
and velocity field with the ratio of the radio and
emission-line sizes, but it is of low statistical significance and
there is a very large dispersion. 
The change in properties at redshifts above $z \sim 0.6$ could reflect
a difference in environments of the host galaxies, with the hosts
inhabiting higher density regions with increasing redshift
(e.g., Hill \& Lilly 1991).  
The mass of ionized gas and the apparent enclosed dynamical mass are
correllated and both increase steeply with redshift and/or radio power.

The origin of the velocities remains uncertain.  The data do not
require jet-gas interactions to explain the kinematics and 
superficially are slightly more consistent with gravitational
origins for the bulk of the kinematics.  If the line width reflects the underlying
gravitational potential, the observed FWHM traces the velocity
dispersion of the host galaxy or its surrounding group or cluster.  
The highest velocities seen then
point to interesting environments for intermediate and high redshift
radio galaxies.  Turbulent interactions with the expanding
radio source as the origin of the kinematics are certainly not ruled out.
In the jet interaction scenario, the maximum velocities
seen in the nebula can be used to constrain the density of the pre-shock gas to be
roughly n$_{\rm e} > 0.6$ cm$^{-3}$.

\end{abstract}

\section    {Introduction}

We report the results of an investigation of the kinematic, ionization
and morphological properties of the emission-line nebulae in powerful
radio galaxies. We have analyzed long slit optical spectroscopic and
narrow band imaging observations of large samples of low (Baum,
Heckman, and van Breugel 1990).  Tadhunter, Fosbury, \& Quinn 1989) and
intermediate to high redshift radio galaxies (McCarthy, Spinrad \& van
Breugel 1995; McCarthy, Baum, and Spinrad 1996). We utilize radio data
from the literature. The combined data allow us to address questions
concerning as the nature and origin of the emission line gas, the
source of energy for the kinematic behavior of the gas, the
relationship of the gas and radio activity to the host galaxy, and the
nature of changes in gaseous environment or host properties as a
function of epoch, environment, radio power and source structure.

Previous such studies have focused primarily on narrow redshift ranges
or specific source types (e.g., Baum, Heckman \& van Breugel 1992;
Tadhunter et al. 1989, Gelderman \& Whittle 1994; Villar-Martin et al.
1998).  The sample amassed here allows us to address the relevant
issues over a much wider range of radio luminosity and redshift.  In
particular, with this sample, we examine changes in the properties of
the nebulae and their relationship to the radio source and host galaxy
over the range of redshifts where we currently believe that both the
host galaxies of powerful radio sources and their environments are
evolving rapidly, $0.3 < z < 2$.  Similarly, we examine the evidence
for increased interaction between the radio source and its gaseous
environment as a function of extended radio power and redshift.

In the low redshift sample considered by Baum, Heckman, \& van Breugel
(1992), they concluded that the bulk of the emission line gas
kinematics were dominated by gravitational motions. For that sample it
proved straightforward to statistically separate gas kinematics
attributable to  radio jet-gas interactions from those determined by
the underlying gravitational potential of the host galaxy.  If we are
able to perform a similar separation at higher redshift, then we could
use the gas kinematics as a function of redshift as a tracer of the
evolution of the gravitational potential of the underlying host
galaxies (the bright ellipticals) and their environments. Conversely,
if we can show that the gas kinematics either in individual sources or
overall, are due to interactions with the out-flowing radio jets, then
carefully follow-up studies of individual sources should lead to a
better understanding of jet physics (e.g., Clark et al. 1998). Lastly,
the apparent increase in the number of close companion galaxies and
possible clusters at $z > 0.5$ (e.g. Hill \& Lilly 1991; Yates et al.
1989; Ellingson, Green, \& Yee 1991) offer the possibility that
kinematics of the emission-line gas may probe tidal interactions and
cluster potentials on 100 kpc scales.

The 3CR and 1Jy class  sources that we consider here comprise a
representative sample of the most luminous radio sources at all
redshifts less than 3.  At redshifts greater than $\sim 0.1$, they are
primarily of the double-lobed Fanaroff \& Riley (1974) type II
morphologies and have luminosities that range from $10^{42} - 10^{45}$
erg sec$^{-1}$. The emission-line regions are seen on scales of up to
300 kpc and arise from low density gas in moderate to high ionization
levels.  The luminosities of the emission-line regions range from
$10^{41} - 10^{44}$ erg sec$^{-1}$ and scale roughly with the radio
luminosity (e.g.  Rawlings and Saunders 1991; Baum, Heckman and van
Breugel 1989a,b; McCarthy 1993; Xu, Livio, and Baum 1999). The total
mass of ionized gas is uncertain but reasonable estimates range up to a
few $\times 10^8$M$_{\odot}$.  The source of the ionization is also
uncertain and is probably a mix of collisional heating and
photo-ionization, with nuclear photo-ionization likely dominating the
total line luminosities over a large range of radio power and redshift
(e.g. Robinson et al. 1987; Baum and Heckman 1989a,b; Rawlings and
Saunders 1991; Baum, Zirbel, and O'Dea 1995; Villar-Martin et al. 1997,
Tadhunter et al. 1998).  The general alignment of the emission-line
regions with the axes defined by the radio lobes in nearly all powerful
Fanaroff and Riley Class II type radio galaxies (e.g. Baum \& Heckman
1989a,b; McCarthy et al. 1987; de Vries et al. 1999) is taken as further
support for the presence of central photo-ionization by an anisotropic
UV source. The precise spatial coincidence between radio hot-spots and
jets with bright emission-line regions in some sources (e.g. de Vries
et al. 1999; Miley et al. 1992, Villar-Martin et al, 1998), and the
correlation between lobe and emission-line asymmetries (McCarthy, van
Breugel, \& Kapahi 1991) argues that shock heating is also an important
process (e.g. Bicknell \& Koekemoer 1995; Best et al.  1999).

In section 2, we describe the definition of the sample, the origin of
the data, and the derived quantities which are used in the analysis.
In section 3, we describe the results of the statistical analysis of
the derived quantities. In section 4, we describe the correlation
results.  In section 5, we discuss the implications of these results
and finally in section 6 we summarize the results and their
implications.

\subsection {The sample }

The majority of the objects are drawn from the 3CR sample (Bennet 1962;
Spinrad et al. 1985, Djorgovksi et al. 1988; Strom et al. 1990). At
redshifts larger than $\sim 0.2$ nearly all of the 3CR galaxies that
are known to have emission lines with extents of more than $5^{''}$ are
included in our sample. The source samples in Baum, Heckman, \& van
Breugel (1990, 1992) \& McCarthy, Baum, \& Spinrad (1996) were drawn
from the emission-line imaging surveys of a 408 MHz radio flux limited
sample of low redshift  equatorial radio sources by Baum et al. (1988)
and a sample of intermediate and high redshift 3CR galaxies imaged by
McCarthy et al. (1995). The latter sample becomes significantly
incomplete for $1.2 < z < 1.6$, where the principal lines that show
large spatial extents (e.g. [OIII]5007, [OII]3727, Ly$\alpha$) are
either unreachable or are in regions of very strong sky emission. In
the range $0.2 < z < 1.2$ our sample contains all of the 3CR galaxies
with large emission-line regions except 3CR 172, 3CR 275, 3CR 341, 3CR
244.1, 3CR 284, 3CR 180.  At larger redshifts we do not have adequate
2-D spectra of the extended gas in 3C 437 ($z = 1.48$), 3C 256 ($z =
1.82$), 3C 326.1 ($z = 1.82$), 3C 454.1 ($z = 1.84$), and 3C 257 ($z =
2.4$).  We have excluded the two Virgo sources (3C 272.1 \& 3C 274) from our
analysis as their linear extents are smaller than the resolution element for
nearly all of the other objects in the sample.
In addition to the 3CR sources, at low redshift our sample
includes three Parkes sources (PKS 0634-206, PKS 0745-191, and PKS
1345+125) from the equatorial sample of Baum et al. (1988).  Similarly,
at $z > 2$ we have added three sources from the 408MHz MRC/1Jy survey
(McCarthy et al. 1996). These three objects are thought to be
representative of the large Ly$\alpha$ emission regions in the most
distant radio galaxies.

\section {   The Data}
We use ground-based long slit spectra and emission line images of 3CR
and 1 Jy class radio galaxies that were obtained at the Kitt Peak and
Cerro Tololo National Observatories, Lick Observatory, and Palomar
Observatory.  The details of the observations and the basic data are
presented in Baum, Heckman, and van Breugel (1990), McCarthy, Spinrad
and van Breugel (1995) and references t herein.  To this data we added
similar data from the literature (e.g. Heckman \etal\ 1989; Tadhunter
\etal\ 1989; Clark et al. 1998) as well as published radio maps
(notably the compilation by Leahy, Bridle, \& Strom (1996) of radio
source maps for the 3CR) to assemble the data used to derive the
correlations presented here.

\subsection {Measured quantities} \begin{itemize}
      \item FWHM -  For each velocity field plotted in either Baum,
      Heckman, \& van Breugel (1990) or McCarthy, Baum, \& Spinrad
(1995) we determined a characteristic extra-nuclear  FWHM. We used the
maximum value of the FWHM, excluding both the nucleus and the regions
where the uncertainties are large. We averaged over three spatial
resolution elements at the location of the maximum FWHM in assigning a
characteristic 
velocity width to the rather low signal-to-noise plots of FWHM vs.
distance that are available for these objects. The tabulated values for
the FWHM and our estimate of the associated uncertainties are given in
columns 11 of Table 1.

The non-zero size of the spatial resolution elements in each spectra
introduce a component of broadening to the lines that is not physical
in origin. This is likely to be important in a number cases for which
there are large velocity gradients (e.g. 3C 330) and for objects at
high redshift where the spatial resolution element ($\sim 1.5"$)
corresponds to scales of $\sim 10$ kpc.

  \item V$_{\rm peak}$ and V$_{max}$.  We determined the maximum
peak-to-peak amplitude of the resolved velocity field (V$_{\rm peak}$)
for each slit position and object combination. As with all of the
measurements we averaged over a few spatial resolution elements and
gave low weight to regions of the slit with large velocity
uncertainties.  Since many of the emission-line regions are distributed
asymmetrically with respect to the nucleus, the peak-to-peak amplitude
is often poorly defined. For this reason we define V$_{max}$ - the
maximum one-sided velocity offset with respect to the nucleus,
regardless of its spatial location. Thus objects with large amplitude
velocity fields, but one-sided nebulae (e.g. 3C 356) are properly
compared with smaller amplitude, but symmetric, objects (e.g. 3C 300).
Several of the objects were observed with more than one telescope and
instrument combination, or in more than one position angle.  For these
we determined single values for the V$_{\rm peak}$ and V$_{max}$
velocity amplitudes by choosing the data set with the highest
resolution and signal to noise ratios. For objects with multiple
position angles we use the angle the revealed the large velocities and
line widths.  In the cases of 3C 458 and 3C 265 for which two spatially
offset slit positions were used, we included separate measurements for
each. V$_{\rm peak}$ and V$_{max}$ are tabulated in columns 6 \& 7 of
Table 1.

     \item R$_{\rm Vmax}$ and D$_{\rm [OII]}$. Several of the derived
quantities depend on both the observed velocities, or line widths, and
a characteristic linear scale. We considered a number of different
choices for this scale length and adopted two. The first is the
isophotal size of the emission line region in kpc (denoted D$_{\rm
[OII]}$ in column 8 of Table 1) as determined from emission-line images
(Baum et al. 1988; McCarthy, Spinrad and van Breugel 1995). These were
measured at a fixed rest-frame surface brightness level corresponding
to f(H$\alpha$)$ = 3.0 \times 10^{-15}$ erg sec$^{-1}$ cm$^{-2}$ per
square arcsecond. Images in [OIII], [OII] or Ly$\alpha$ were converted
to equivalent H$\alpha$ surface brightnesses (McCarthy et al. 1995).
The depth reached by the slit spectra and the narrow-band imaging were
not always identical, particularly when considering that not all of the
region of detected emission in the spectra were of sufficient quality
to allow for accurate velocity measurements. We also measured the
distance from the nucleus to the point of maximum velocity directly
from the spectra. This distance, R$_{\rm Vmax}$ is listed in column 9
in Table 1 and is used in several of the plots and computations.

\item{} Radio source sizes, D$_{\rm Rad}$ and l$_{\rm Rad}$ - The
linear (lobe separation) sizes of the radio sources are given as D$_{\rm
rad}$ in Column 12 of Table 1. The angular sizes were either measured
directly from the maps in Baum et al. (1988) or taken from the
compilation in McCarthy, van Breugel and Kapahi (1991), and were then
converted to kpc using H$_0 = 50$ km sec$^{-1}$ Mpc$^{-1}$ and q$_0 = 0.1$.
As discussed above, the emission-line regions are
often one sided and a simple comparison between the maximum size of the
radio source and emission-line region may not give a proper reflection
of the degree of spatial coincidence between the radio and
line-emitting material. We used the most recent maps to determine the
angular size of the radio sources on the same side of the nucleus as
the gas whose velocity field we measure. Thus for sources with
asymmetric emission-line regions we tabulate the length of one arm of
the radio source, while for more symmetric nebulae we tabulate the
total size of the source.  This emission-nebula dependent measure of
the radio source extent is given as l$_{\rm rad}$ in Column 13 of
Table 1.  Sources with one sided nebulae and close coincidences between
radio lobes and emission line regions (e.g. 3C 441) can now be
recognized by the rough equalities of the two sizes in columns 8 and
13, in Table 1.

\end{itemize}

\subsection {Derived quantities}

\begin{itemize}

\item Gas mass, M$_{\rm gas}$ (column 14, Table 1).  We estimated the
total mass of line-emitting gas for each object in our sample. The
masses were derived from the measured luminosities and characteristic
scales determined from the emission-line images, and an assumed value
of the ionization parameter of 1\%. The filling factor, in units of
$10^{-5}$, is given by, $f_{v5} = 41.3 \times (U_{-2}\times
r_{10}/L_{44})^{0.5} $, where $U_{-2}$ is the ionization parameter in units of
$10^{-2}$, $U =
Q(ion)/4\pi r^2n_ec$. The mass of gas, in units of $10^8$M$_{\odot}$,
is $ 1.4 \times (L_{44}\times_{10}^3\times f_{v5})^{0.5}$, where
$L_{44}$ is the equivalent [OII]3727 luminosity in units of $10^{44}$
erg/sec and $r_{10}$ is the size of the emission-line nebulae in units
of 10 kpc. These masses are highly uncertain as they, 1) make very
simplified assumptions about the geometry of the line-emitting gas (a
sphere of radius $r_{10}$), 2) adopt the same value of $U_{-2}$ for all of
the sources and 3) make no correction for ionization fraction,
abundances etc.

\item
     Dynamical Mass, M$_{\rm dyn}$ (column 17, Table 1).  To the extent
that the large scale velocity fields reflect the gravitational
potential and are representative of the circular velocity, they can be
used to infer an interior dynamical mass. We use the distance to the
peak emission-line velocity, R$_{\rm Vmax}$ and the maximum velocity to
compute the inferred enclosed mass in units of $10^8$M$_{\odot}$ as
$0.0022 \times $R$_{\rm Vmax} \times$ V$_{\rm max}^2 $ with $R_{\rm
Vmax}$ in kpc and V in km/sec.  This apparent or inferred dynamical
mass only really measures mass if the kinematics are
dominated by rotation in response to the gravitational field, and will
have an entirely different meaning if the gas kinematics are set via
radio source - cloud interactions.

\end{itemize}

\section {Results }

     Our sample and the derived data are listed in Table 1. For each
     source we list a single set of measurements corresponding to the
slit position angle with the largest velocities. For 3C 458 and 3C 265
we list two sets of measurements as they were taken at two different
slit positions, one centered on the nucleus and a parallel slit
position offset from the nucleus. For a few objects we are not able to
measure meaningful values for all of the quantities and these are left
blank in the table.

In Figures 1-4  we plot various relations between the observed and
derived parameters described above.  In Table 2 we give the derived
correlation coefficients corresponding to plots shown in Figures 1 - 4.

{\tiny
\begin{planotable}{lrllrrrrrrrrrrrr}
\tablewidth{0pt}
\tablecaption{Data Table}
\tablehead
{
\colhead{Source}      &         
\colhead{FR}          &         
\colhead{PA}          &         
\colhead{$z$}         &         
\colhead{P$_{\rm{408}}$} &         
\colhead{V$_{\rm{peak}}$} &         
\colhead{V$_{\rm{max}}$}   &         
\colhead{D$_{\rm{[OII]}}$}    &         
\colhead{R$_{\rm{Vmax}}$} &         
\colhead{L$_{\rm{[OII]}}$}    &         
\colhead{FWHM}        &         
\colhead{D$_{\rm{Rad}}$}    &         
\colhead{l$_{\rm{Rad}}$}    &         
\colhead{M$_{\rm{gas}}$}     &         
\colhead{M${\rm{dyn}}$}      &         
\colhead{Ref}         \cr         
\colhead{1}      &         
\colhead{2}          &         
\colhead{3}          &         
\colhead{4}         &         
\colhead{5} &         
\colhead{6} &         
\colhead{7}   &         
\colhead{8}   &         
\colhead{9} &         
\colhead{10}    &          
\colhead{11}      &         
\colhead{12}      &         
\colhead{13}    &         
\colhead{14}    &         
\colhead{15}     &         
\colhead{16}    \cr           
}

\startdata
3CR 272.1 &  I & 085 &0.003 & 23.7 &250  &125  & \ \ \   2  & \ \ \ 0.6 & \ \ 39.1 & 150 $\pm$ 050 &  14 & 14 & 4.6 &  9.3 & 1 \cr
3CR 274.0 &  I & 000 &0.004 & 25.6 &370  &250  &  21  &  0.8 &40.2 & 219 $\pm$ 100 &  52 & 52 & 6.7 &  ... & 1 \cr
3CR 264.0 &  I & 085 &0.021 & 25.4 &17   &8.5  &   4  &  0.3 &39.8 & 450 $\pm$ 150 &  29 & 15 & 4.2 &  6.7 & 1 \cr
3CR 442.0 &  I & 126 &0.026 & 25.6 &140  &70   &   7  &  2   &40.2 & 300 $\pm$ 100 & 385 &192 & 5.8 &  9.3 & 1 \cr
3CR 078.0 &  I & 120 &0.029 & 25.7 &30   &15   &   7  &  1.6 &40.4 & 480 $\pm$ 050 & 141 & 71 & 5.7 &  7.9 & 1 \cr
3CR 088.0 &  I & 150 &0.030 & 25.6 &60   &40   &   8  &  3   &40.5 & 350 $\pm$ 100 & 189 & 95 & 6.1 &  9.0 & 1 \cr
3CR 353.0 &  I & 160 &0.030 & 26.7 &100  &100  &   6  &  2   &40.2 & 400 $\pm$ 200 & 223 &112 & 5.8 &  9.7 & 1 \cr
3CR 098.0 &  I & 163 &0.031 & 26.0 &310  &210  &  23  &  8   &40.6 & 250 $\pm$ 050 & 264 &133 & 6.9 & 10.9 & 1 \cr
PKS 0634  & II & 124 &0.056 & 26.5 &480  &260  &  60  & 14   &42.2 & 150 $\pm$ 100 &1169 &117 & 0.0 & 11.3 & 1 \cr
3CR 405.0 & II & 160 &0.056 & 26.5 &105  &62   &  13  &  3   &41.9 & 350 $\pm$ 050 & 187 & 94 & 6.5 &  9.4 & 1 \cr
3CR 403.0 & II & 020 &0.059 & 26.4 &480  &260  &  16  &  6   &40.7 & 200 $\pm$ 100 & 333 &167 & 6.7 & 11.0 & 1 \cr
3CR 033.0 & II & 019 &0.059 & 26.7 &310  &220  &  17  &  4   &41.4 & 300 $\pm$ 150 & 419 &211 & 6.6 & 10.6 & 1 \cr
3CR 192.0 & II & 145 &0.060 & 26.3 &340  &230  &  39  & 20   &41.1 & 150 $\pm$ 100 & 305 &191 & 7.8 & 11.4 & 1 \cr
3CR 285.0 & II & 140 &0.079 & 26.2 &390  &260  &  21  &  5   &41.2 & 200 $\pm$ 100 & 374 &187 & 6.7 & 10.9 & 1 \cr
3CR 227.0 & II & 121 &0.086 & 26.7 &260  &160  & 144  & 20   &42.0 & 250 $\pm$ 100 & 501 &274 & 8.0 & 11.1 & 1 \cr
3CR 433.0 & II & 129 &0.102 & 27.1 &410  &410  &  31  &  8   &40.6 & 500 $\pm$ 200 & 161 & 82 & 6.9 & 11.5 & 1,2 \cr
PKS 0745  &  I & 112 &0.103 & 26.7 &75   &75   &  34  &  7   &42.5 & 355 $\pm$ 100 &  31 & 16 & 8.0 &  ... & 1 \cr
PKS 1345  &  0 & 060 &0.122 & 26.8 &165  &165  &  26  &  6   &42.2 & 550 $\pm$ 200 & 0.3 & 0.3 & 7.1 & 10.5 & 1 \cr
3CR 381.0 & II & 155 &0.161 & 27.0 &480  &280  & 127  & 29   &42.3 & 200 $\pm$ 200 & 257 &130 & 8.3 & 11.7 & 2 \cr
3CR 063.0 & II & 090 &0.175 & 27.2 &290  &280  &  63  & 10   &41.9 & 500 $\pm$ 100 &  88 & 44 & 7.4 & 11.2 & 1 \cr
3CR 033.1 & II & 054 &0.181 & 27.1 &150  &100  &  58  & 16   &41.8 & 300 $\pm$ 200 & 884 &327 & 7.7 & 10.6 & 2 \cr
3CR 196.1 &  I & 052 &0.198 & 27.2 &50   &50   &  25  &  7.0 &41.5 & 400 $\pm$ 100 &  17 & 17 & 7.0 &  9.6 & 1 \cr
3CR 079.0 & II & 322 &0.256 & 27.6 &530  &350  & 117  & 53   &42.4 & 200 $\pm$ 200 & 454 &195 & 8.8 & 12.2 & 2 \cr
3CR 460.0 & II & 215 &0.268 & 27.1 &220  &120  &  38  & 16   &41.6 & 500 $\pm$ 250 &  32 &  8 & 7.7 & 10.7 & 2 \cr
3CR 300.0 & II & 000 &0.270 & 27.6 &450  &380  &  72  & 54   &42.7 & 300 $\pm$ 100 & 526 &164 & 8.9 & 12.2 & 2 \cr
3CR 458.0 & II & 040 &0.290 & 27.5 &280  &180  & 232  & 68   &42.4 & 325 $\pm$ 100 &1099 &633 & 9.0 & 11.7 & 2 \cr
3CR 458.0 & II & 040 &0.290 & 27.5 &350  &300  & 232  & 68   &42.4 & 350 $\pm$ 150 &1099 &633 & 9.0 &  ... & 2 \cr
3CR 299.0 & II & 000 &0.367 & 27.7 &320  &320  &  88  & 39   &42.5 & 600 $\pm$ 100 &  78 & 20 & 8.6 & 11.9 & 2 \cr
3CR 306.1 & II & 180 &0.441 & 27.9 &080  &060  &  51  & 15   &42.7 & 400 $\pm$ 200 & 690 &350 & 7.9 & 10.1 & 2 \cr
3CR 435.0 & II & 229 &0.471 & 27.8 &250  &150  & 281  & 76   &42.7 & 300 $\pm$ 100 &  94 & 94 & 9.1 & 11.6 & 2 \cr
3CR 330.0 & II & 230 &0.550 & 28.5 &650  &450  &  85  & 82   &43.5 & 450 $\pm$ 200 & 517 &259 & 9.4 & 12.6 & 2 \cr
3CR 169.1 & II & 126 &0.633 & 27.9 &1500 &1400 &  73  & 61   &43.0 & 500 $\pm$ 100 & 339 &178 & 9.1 & 13.4 & 2 \cr
3CR 337.0 & II & 090 &0.635 & 28.2 &320  &260  & 105  &132   &41.9 & 400 $\pm$ 200 & 384 &161 & 9.4 & 12.3 & 2 \cr
3CR 034.0 & II & 000 &0.689 & 28.3 &150  &150  & 170  & 28   &43.9 &           ... & 425 &425 & 8.7 & 11.1 & 2 \cr
3CR 441.0 & II & 144 &0.707 & 28.4 &1100 &1050 & 162  &138   &42.7 & 100 $\pm$ 100 & 308 &112 & 9.6 & 13.5 & 2 \cr
3CR 277.2 & II & 055 &0.766 & 28.4 &950  &950  & 316  &131   &43.5 & 500 $\pm$ 100 & 521 &193 & 9.8 & 13.4 & 2 \cr
3CR 340.0 & II & 090 &0.775 & 28.4 &110  &060  &  38  & 19   &42.9 & 450 $\pm$ 100 & 436 &223 & 8.1 & 10.2 & 2 \cr
3CR 352.0 & II & 150 &0.806 & 28.5 &650  &350  &  60  & 58   &43.3 & 850 $\pm$ 100 & 100 &100 & 9.1 & 12.2 & 2 \cr
3CR 265.0 & II & 115 &0.811 & 28.7 &680  &450  & 283  &105   &44.1 & 500 $\pm$ 100 & 768&768 & 9.8 & 12.7 & 2 \cr
3CR 265.0 & II & 115 &0.811 & 28.7 &500  &500  & 283  &105   &44.1 & 500 $\pm$ 100 & 768&768 & 9.8 & 12.8 & 2 \cr
3CR 280.0 & II & 090 &0.998 & 29.0 &550  &500  &  98  & 72   &44.0 & 700 $\pm$ 100 & 135& 68 & 9.4 & 12.6 & 2 \cr
3CR 356.0 & II & 148 &1.079 & 28.8 &1500 &1500 &  85  & 73   &43.9 & 700 $\pm$ 100 & 302&302 & 9.4 & 13.6 & 2 \cr
3CR 368.0 & II & 065 &1.132 & 28.9 &575  &550  &  91  & 43   &44.1 & 1400$\pm$ 100 &  93& 93 & 9.0 & 12.4 & 2 \cr
3CR 267.0 & II & 090 &1.140 & 28.9 &380  &380  &  31  & 21   &43.3 & 750 $\pm$ 250 & 416&192 & 8.3 & 11.8 & 2 \cr
3CR 324.0 & II & 090 &1.206 & 29.0 &725  &500  &  92  & 53   &44.0 & 1000$\pm$ 200 & 107&107 & 9.2 & 12.5 & 2 \cr
3CR 266.0 & II & 180 &1.275 & 29.0 &550  &450  &  47  & 33   &43.9 & 900 $\pm$ 200 &  45& 45 & 8.8 & 12.2 & 2 \cr
3CR 266.0 & II & 180 &1.275 & 29.0 &1350 &1350 &  47  &  0.1 &43.9 &           ... &  45& 45 & 0.0 &  ... & 2 \cr
3CR 294.0 & II & 180 &1.786 & 29.4 &1300 &1200 & 152  & 91   &44.4 & 1100$\pm$ 200 & 173& 77 & 9.7 & 13.4 & 2 \cr
MRC 0457  & II & 065 &1.960 & 28.9 &600  &550  & 114  & 67   &44.4 & 1400$\pm$ 300 & 200& 96 & 9.5 & 12.6 & 2 \cr
MRC 0406  & II & 128 &2.44  & 29.8 &1900 &1000 &  76  & 45   &44.5 & 1000$\pm$ 300 &  89& 55 & 9.2 & 14.0 & 2 \cr
MRC 2104  & II & 021 &2.49  & 29.6 &490  &350  & 139  & 23   &44.4 & 800 $\pm$ 200 & 269&134 & 8.6 & 11.8 & 2 \cr
\end{planotable}
}
\noindent {\bf Notes to Table 1:} Velocities are in
units of km sec${-1}$, radio power is in units of erg sec$^{-1}$ Hz$^{-1}$,
emission-line luminosities are in units of erg sec$^{-1}$, all length scales
are in kpc, M$_{gass}$ and M$_{dyn}$ are in solar mass units. Luminosities
and masses are shown as logarithms. 

\noindent references: 1) Baum, Heckman \& van Breugel (1990); 2) McCarthy, Baum, \& 
Spinrad (1995).


\begin{table}[htb]
\begin{center}
\begin{tabular}{ll}
\multicolumn{2}{c} {Fits to Correlations }  \\
\hline
Relationship &  Correlation \cr
             & Coefficient \cr
(1) & (2)   \cr
\hline
P$_{\rm 408}$ vs $z$ &  0.98   \cr
R$_{\rm Vmax}$  vs $z$&  0.82   \cr
M$_{\rm gas}$  vs $z$ &  0.77   \cr
R$_{\rm Vmax}$  vs V$_{\rm max}$ &  0.76   \cr
V$_{\rm max}$ vs M$_{\rm gas}$ &  0.76   \cr
M$_{\rm dyn}$  vs $z$&  0.75   \cr
P$_{\rm 408}$ vs V$_{\rm max}$ &  0.67   \cr
V$_{\rm max}$  vs $z$ &  0.64   \cr
M$_{\rm gas}$ vs  M$_{\rm dyn}$ &  0.64   \cr
P$_{\rm 408}$ vs FWHM &  0.63   \cr
FWHM  vs $z$&  0.62   \cr
l$_{\rm Rad}/$R$_{\rm Vmax}$ vs V$_{\rm max}$ & 0.43$^{*}$   \cr
l$_{\rm Rad}/$R$_{\rm Vmax}$ vs FWHM & 0.41$^{*}$   \cr
D$_{\rm Rad}$ vs D$_{\rm [OII]}$ &  0.34   \cr
FWHM vs M$_{\rm gas}$ &  0.32   \cr
D$_{\rm rad}$ vs FWHM  & 0.26$^{*}$   \cr
D$_{\rm rad}$ vs V$_{\rm max}$ &  0.26   \cr
V$_{\rm max}$ vs FWHM &  0.20   \cr
D$_{\rm rad}$  vs $z$&  0.10   \cr
\hline
\end{tabular}
\end{center}
\caption{The results of a least squares linear fit to the relationships in
log-log space. Uncertainties are in parentheses. }
\end{table}
~

\section {Correlation Results}

Below we describe the most relevant correlation results.

\subsection {Radio Power with Redshift} In our sample, as in most flux limited samples 
extending to significant redshifts, radio power and redshift are strongly correlated
making the disentanglement of redshift versus radio power effects a
difficult challenge.

\subsection{Log Gas Mass versus log Redshift}

There is an apparent correlation of the ionized gas mass with redshift,
with ionized gas mass increasing roughly linearly (in a log-log plot)
with redshift (Figure 1a). Further, the plot shows that the FR1 and FR2 sources
follow the same correlation.  If this correlation is real (and not just
a selection effect - see below), then there are two straightforward
possible explanations for it. \begin{itemize}

\item The mass of cold gas within the parent galaxies increases with
redshift, such that cold gas was more abundant in radio galaxy hosts at
earlier epochs. This is consistent with (1)  increased bending in radio
sources at high redshift (e.g., Barthel \& Miley 1988); (2) the higher
fraction of compact GHz Peaked Spectrum quasars at high redshift (e.g.,
O'Dea 1998); (3) the alignment effect occurring preferentially at high
redshift (e.g., McCarthy 1993).

\item The flux of ionizing radiation, and hence the measured mass of
ionized gas for radiation bounded nebulae, increases with radio power
(redshift).  \end{itemize}

However, some caution is needed in interpreting the data, as the
correlation may owe at least in part, to an observational selection
effect, since sources with small masses of ionized gas at high redshift
would not have been detected as extended emission line sources and so
would not have been included in follow up spectroscopic studies such as
this one. That is, the limiting surface brightness curve defines an
excluded region of the plot.  However, despite this caveat, the sources
in our sample do nevertheless show an order of magnitude or more
variation in observed gas mass such that at low redshifts, (or radio
powers) we do not find sources with as high gas masses as we do at
higher redshift (power).

\subsection{ Dynamical Mass versus $z$}

We also find an apparent correlation of redshift (and radio power) and
enclosed dynamical mass (Figure 1b).  FR1s have systematically low dynamical masses
at a given redshift, and we find that the highest dynamical mass
objects are double kinematic systems (see the correlation of redshift
with V$_{max}$ below), which are found preferentially at the highest z.  The
maximum derived dynamical masses are comparable to the masses of
clusters, as derived from x-ray and lensing studies (Squires
\etal\ 1997; Allen, Fabian, \& Kneib 1996; Miralda-Escud\'e \& Babul
(1995)), at comparable distances from the cluster center (e.g., 100
kpc). There are two ways to understand this correlation, assuming for
the moment that the velocities do measure the gravitational potential
and not the radio source jet thrust.

\begin{itemize}

\item The mass of the host galaxy is larger at redshifts and radio
powers. This seems somewhat unlikely given that radio galaxies live in
the most massive galaxies even at low redshifts, and that it is well
known (Ledlow \& Owen 1996; de Vries et al 1998) that at present 
FR1s (and not FR2s) on average occupy higher mass
systems at a fix radio luminosity.

\item The ionized gas probes progressively larger scales at higher
redshift, thereby enclosing greater and greater dynamical mass. This
could be due either to the presence of more cold gas in the host
galaxies at higher z (see above) or, to the increased UV ionizing flux
put out by the more powerful radio galaxies at high redshift in our
sample.  \end{itemize}

\subsection{ Dynamical mass versus M$_{gas}$}
There is a strong apparent correlation of inferred dynamical mass with gas mass,
with a slope near one half - that is dynamical mass proportional to the
square root of the gas mass (Figure 2a).  The correlation for FR1 and FR2 radio
galaxies is the same, with the FR1s occupying the low mass portion of
the relationship.

If the correlation is real (and the inferred dynamical mass is really
a measure of enclosed mass) it suggests that \begin{itemize}

\item there is a constant fraction of gas to gravitational mass of
$\sim 0.1\%$ and \item the ionized gas mass measures gas mass and not
ionizing flux.  \end{itemize}

We tested for the possibility that this correlation is artificial, since
one of the quantities depends on R and the other on R$^3$ by plotting
$R\times V$ against the [OII] luminosity directly; 
the correlation, though noisier, remains (see Figure 2c).

\subsection { V$_{max}$ vs. R$_{V_{max}}$ }

There is an apparent correlation of maximum velocity versus the radius
in kpc at which that maximum velocity is measured, with a slope near
unity, such that the larger the distance,  the larger the velocity (Figure 2d).
FR1s fit on the same relationship as FR2s.  This can be most
simply interpreted in the following way.  The velocities are
gravitationally induced, all radio galaxies have basically the same
intrinsic masses and environments, and the further out one probes 
the gas velocity, the greater the velocity one measures since a larger
mass is encircled.  In the most powerful radio sources (or the highest
redshift sources) - the larger luminosity or ionizing radiation (which
is assumed to scale with radio luminosity) illuminates gas
at larger radii from the galaxy nucleus, producing this result.

Interestingly, we see from this plot that the maximum velocity which is
measured is roughly $1500$ km sec$^{-1}$. What can this mean?  Two
alternate scenarios present themselves; \begin{itemize}

\item If the velocities are gravitational in origin, than as mentioned
above, 1500 km sec$^{-1}$ is roughly the velocity predicted for gas in
equilibrium within a very rich cluster at a distance of $\sim 100$ kpc
from the nucleus (e.g., Mazure \etal\ 1996).  It appears, however, that
many of the v $> 800$ km sec$^{-1}$ systems are composites (i.e., two
galaxies); the maximum  velocity would then be a measure of the peak
galaxy encounter velocity. Again, 1500 km sec$^{-1}$ is a sensible pair-wise 
encounter velocity.
It is interesting to consider whether pairs can produce such
velocities without obvious clusters at redshifts of $0.6 - 0.7$ (e.g.
3C 169.1, 3C 277.2 where these high velocities are also seen).  We
return to this later in the discussion.

\item If the velocities are shock induced from the radio source and not
gravitational, then the maximum of 1500 km sec$^{-1}$ may correspond
to the maximum velocity achieved before the gas is heated to
temperatures such that the radiative cooling time is longer than
the source lifetime. Bicknell, Dopita and O'Dea (1997)
assert that detailed fast radiative shock models fit the following
relationship:  $t_{rad,6} \sim  1.9 n^{-1} V_3^{2.9}$ where $V_3$ is
the shock velocity in units of 1000 km/sec, n is the preshock density.
Thus for V = 1500 km/sec and a radio
lifetime = $10^7$ years, only clouds with densities
greater than $0.6$ cm$^{-3}$ will produce significant optical and near-UV line emission.
At higher
shock velocities or lower preshock densities the heated gas will
be too hot and diffuse to cool within the radio source lifetime.
\end{itemize}

\subsection {FWHM and V$_{max}$ versus l$_{\rm Rad}$/R$_{\rm Vmax}$}

The plots of FWHM and V$_{max}$ versus l$_{\rm Rad}$/R$_{\rm Vmax}$ 
(Figures 4a and 4b) are helpful in disentangling the relationship
between the radio source and the observed velocities; if the radio
source is responsible for inducing the turbulence in the gas then we
might expect that there would be a strong correlation  between the FWHM
(or V$_{max}$) and the ratio of radio source to emission line nebulae size.  That is,
we might expect that when the radio source and the nebulae are
comparably sized, the gas will be be most kinematically disturbed.

There appears to be a weak correlation with a trend towards
larger line widths in objects with comparable radio and emission-line sizes. 
However, this is not a statistically robust
correlation and there is a large ranges in FWHM and V$_{max}$ values for a
given relative size.  Interestingly, the FR1s follow the same
correlation as the FR2s. Longair et al. (1999) report a significant correlation between
source size and off-nuclear line widths in a small sample of $z \sim 1$ 3CR galaxies.

\subsection{ FWHM versus $z$} There is an apparent correlation of line width and
redshift (and radio power), which warrants further examination.
The plot of FWHM versus $z$ (Figure 3a) is essentially flat for $0.01 < z < 0.6$
and rises steeply (or jumps) for larger
redshifts. The highest line widths are seen at z$>1$; for $z < 1$
the maximum line width is 500 km sec$^{-1}$ and the mean is $370 \pm 150$ km
sec$^{-1}$ while for $z > 1$, the maximum is 1500 km
sec$^{-1}$  and the mean is $1000 \pm 250$ km sec$^{-1}$.  
The FR1 sources have relatively high mean line widths relative to FR2 sources
at the same redshift with an FR1 mean of $\sim 385 \pm 95$ km sec$^{-1}$ and
an FR2 mean of $\sim 285 \pm 130$ at z$<0.2$.

The change in properties at redshifts  above $z \sim 0.6$ could signal
a difference in environments of the host galaxies.
At low redshift FR2s are predominantly found in the field or
groups of galaxies (e.g., Prestage \& Peacock 1988 ), while at higher
redshifts they are believed to be found in environments with densities reaching
those of  rich clusters
(e.g., Hill \& Lilly 1991).  Thus, if the line width reflects the underlying
gravitational potential, the observed FWHM may trace the velocity
dispersion of the host galaxy or its surrounding group or cluster.  The
systematically high FWHM of FR1 sources relative to FR2 sources at low
redshifts is grossly consistent with this scenario, since FR1s
occupy more massive galaxies at a given radio luminosity
and inhabit denser environments at low redshifts than do FR2s (Ledlow
\& Owen 1996; Hill \& Lilly 1991).  

In the alternative scenario, where the FWHM reflects an interaction
with the radio source, this same plot can be interpreted as FWHM versus
radio power. At higher powers the interaction is stronger and the
turbulence induced in the gas greater. In is not clear how this scenario 
would produce the sudden change in line widths at $z > 0.6$.

\subsection{ V$_{max}$ versus z} The V$_{max}$ versus $z$
plot is similar to the FWHM versus $z$ plot and could reflect the same
process involved in the change in line widths at large redshifts. 
Specifically, if one excludes the FR1s, which have
systematically low velocities at a given redshift (see the subsection 4.9
below for an explanation of the FR1
behavior) then the velocities are flat for $0.01 < z < 0.7$ 
at a mean of $\sim 230 \pm 240$ km sec$^{-1}$ and then the mean value jumps
dramatically to $v \sim 720 \pm 400$ km sec$^{-1}$ for $z > 1$.
This is, again, most likely due to a change in environments of the host
galaxies, but may also reflect the importance of radio power in
determining the overall gas kinematics.  In the latter scenario, it is,
however, difficult to understand why one would see a `jump' at
intermediate redshifts as opposed to a continuous correlation of
velocity with radio power.

Special consideration should be given to the highest V$_{max}$ systems. We
note that of the sources with V$_{max}$ greater than $\sim 800$ km
sec$^{-1}$ approximately six of these sources show double kinematic
structure. 3C 169.1, 3C 356 and MRC 2104-242 for example, are systems
with two fully detached emission line regions with velocity differences
of several hundred km sec$^{-1}$ or more (see McCarthy, Baum, \&
Spinrad (1996) for velocity versus distance plots).  These are likely
to be instances where we are seeing gas in two distinct kinematic
systems, for example two independent galaxies, or gas in "split-line"
systems where the gas has been wrapped around two sides of expanding
radio plasma interacting with out-flowing radio plasma (e.g. Clark et
al. 1997; Capetti et al. 1999) We also note that the maximum values of V$_{\rm max}$
seen are roughly $\sim 1200-1500$ km sec$^{-1}$. See also the
discussion under 4.5 above.

\subsection { FWHM/V$_{max}$ versus P$_{408}$}

The FR1 sources cleanly separate from the FR2 sources on this plot of FWHM/V$_{max}$ versus P$_{408}$
(Figure 3d).
In the mean the FR1s have appreciably higher ratios of FWHM to
maximum velocity than the FR2s (8:2) at the same radio power. As Figures 
3c and 3d show, FR1s have both larger line widths and lower maximum
velocities at a given redshift than do the FR2s. This is consistent
with the earlier results from Baum \etal\ (1992)  that indicated that
the emission line gas in FR1 sources was dominated by turbulent
motions, while that in FR2s showed a stronger component of systematic
motions (e.g., rotation, outflow, or inflow).  This plot also shows
that the ratio of turbulent to systematic motion shows no evolution
with radio power or redshift for the FR2s. This seems to suggest a
common origin for both the turbulent and the systematic motions seen in
the gas.

\section {Discussion}

As is typically the case for flux limited samples, radio power and
redshift are very tightly coupled in our sample, making the deconvolution
of radio power and redshift affects difficult, at best.
A crucial question we must, nevertheless,
seek to address is whether the kinematics
we observe are due primarily to interactions between the gas and
the outflow from the central engine  or whether they reflect the
underlying potential of the host galaxy.  A similar question was addressed
before for the emission line gas in Seyfert galaxies (Whittle 1992) and
in nearby FR1 and FR2 radio galaxies (Baum \etal\ 1992).  In both cases
the conclusion was that the bulk of the velocities observed were dictated
by the underlying gravitational potential and not by jet-gas interactions,
though in both cases examples where that rule was violated and jet-gas
interactions were dominant, were noted. In the
high redshift/high radio power sample we address in this paper where
we expect the outflow to be stronger and the gas environment to be
richer, we might expect jet-gas interactions to be more dominant.
Looking at the statistical data as a whole, however we find no evidence which
{\it requires} jet interaction as the source of the gas kinematics.
As is the case for their nearer by AGN brethren, the emission line kinematics
even in the bulk of these high redshift/high power sources can be explained
by gravity and not interaction, though as described below if the
kinematics are gravitational in origin, the highest velocities seen
point to interesting environments for intermediate and high redshift
radio galaxies.

The apparent correlations of radio power
with gas FWHM, V${\rm peak}$ and R$_{\rm Vmax}$ point towards the importance
of jet-gas interactions. However, it is not possible to distinguish
with our sample, whether these correlations are primarily with redshift
rather than radio power. Further, additional correlations
with radio power or source size might have been expected if jet-gas
interactions dominate the kinematics of the gas.
For instance, if jet-gas interactions are dominant,
one might have naively expected that either smaller radio sources, i.e., those
which are trapped within the ISM of the host galaxy, or sources in which
the radio source and emission line nebulae were of localized on the same
size scales would evidence stronger jet-gas interactions.
However, no correlations are seen
between the size of the radio source and the FWHM or maximum velocity
of the gas. Similarly, no substantial correlation  is seen between
the FWHM or the maximum velocity of the radio source and the ratio of
the radio to nebular size. Lastly, our sample shows no correlation of radio size with
radio power or redshift, or of radio size with nebular size,
but does show a substantial correlation of
nebular size with redshift and/or radio power. Taken as a whole these
findings would seem to indicate that the properties of the emission line
nebula are coupled with redshift or radio power but not with the
properties of the radio source.

Thus, the kinematics and correlations we see,
can {\it most easily} (though not
necessarily correctly of course) be interpreted in terms of a gravitational
origin for the kinematic properties of the gas.
To wit; we find that the FWHM, maximum velocity, radius at which
maximum velocity is seen, apparent dynamic mass and gas mass all correlate
with redshift (or radio power). In addition, we find that the
maximum velocity and radius of maximum velocity are themselves
correlated, as are the inferred dynamical mass and the gas mass,
and the maximum velocity and gas mass.
A plausible scenario which encapsulates the observations follows.

\begin{itemize}
\item
The emission line nebulae are largely ionization (and not matter) bounded.
That is, there is an abundance of cold gas in the ISM of the radio galaxies
and when there is more ionizing radiation, more of the cold gas, and
to further distances in the ISM, is illuminated and heated.

\item
The UV ionizing flux of the central engine is correlated with radio
power (hence redshift in our sample) such that at higher power (redshift)
there is a higher UV ionizing flux and hence emission line gas is found
to greater radii.

\item The emission line gas
kinematics (both the line widths and the absolute
velocities) are dominated by gravitational effects and therefor reflect the
underlying potential at the distance where the gas is found.

\end {itemize}

If we take this model at face value, for the moment, we can then use it to understand the
nature of the environments of powerful radio galaxies as a function
of redshift. Specifically, at high redshift, where the highest
rotational velocities are seen, we would have to be seeing the gravitational
potential of not only the radio galaxies, but a cluster or group
which surrounded it. The velocity separations seen (1200-1500
km sec$^{-1}$)
approach those of the richest clusters seen at lower redshift (e.g., Mazure
\etal\ 1996). This either indicates that high redshift high power radio
galaxies are uniquely situated in the deepest potentials or it suggests that
the host galaxies of powerful radio galaxies at high redshift are
in pre-virialized environments in which infall may still dominate
the kinematics.

Completely alternative models
in which the kinematics are dominated by interactions with the
radio source cannot be ruled out.
Detailed studies show that such interactions are clearly occurring in some
sources and so must play at least a part in determining some of the kinematics
we see. For instance, an
alternative interpretation of the sources which exhibit two very high velocity
emission systems is that we are observing split line systems created by
interactions of the cocoon of the expanding radio source with cold gas in
the galaxy atmosphere, such as appears to be the case in the low velocity
systems studied in detail such as Cygnus-A (3C405; Tadhunter et al. 1999) and 3C171 (Clark et al. 1998).
If the velocities are shock induced from the radio source and not
gravitational, then the maximum of 1500 km sec$^{-1}$ must correspond
to the maximum velocity achievable without heating the gas to
temperatures so hot that the radiative cooling time is greater than
the lifetime of the radio source. Applying the models of Bicknell, Dopita,
and O'Dea (1998), this
would constrain the pre-shock particle densities of the shocked clouds
we see to be n$_e >  0.6$ cm$^{-3}$.

\section{Summary}

We report the results of a study of the kinematics and morphology
of the emission line nebula in powerful 3CR radio galaxies from redshift
$\sim$ zero to $\sim 4$ based on the data presented in
Baum, Heckman and van Breugel (1990) and McCarthy, Baum, and Spinrad (1996)
and additional data gathered on these and similar sources
from the more general literature.
The 3CR is a well-defined radio-flux limited sample; radio power
(AGN luminosity) is strongly correlated with redshift in this sample.

We investigate the correlation of the kinematic
properties of the gas with other inherent properties, such as
redshift, radio and line luminosity, radio and emission line morphology
and extent. We also investigate correlations with
derived properties such as inferred mass of emission line gas and
apparent dynamical mass, where the apparent dynamical mass is defined to be
the mass required
to gravitationally induce the observed kinematics.
We find that both the inferred mass in emission line gas and the
apparent dynamical mass appear to increase with redshift (or radio power).

For each source we define a maximum emission line velocity (V$_{\rm max}$, the maximum
difference between gas velocity and the
systemic velocity of the galaxy, measured as the emission line velocity
at the continuum peak) and the projected radius at which that maximum velocity
occurs (R$_{\rm Vmax}$).  We find that across the full sample, the
maximum emission line velocity increases with R$_{\rm Vmax}$ and with
the absolute size of the nebula.
We also find that the sources with the highest maximum velocities (1000-1500
km sec$^{-1}$) are those with two independent kinematic systems present.

No correlations are seen
between the size of the radio source and the FWHM or maximum velocity
of the gas. Similar to the above, no substantial correlation  is seen between
the FWHM or the maximum velocity of the radio source and the ratio of
the radio to nebular size.
Lastly, our sample shows no correlation of radio size with
radio power or redshift, or of radio size with nebular size,
but does show a substantial correlation of
nebular size with redshift and/or radio power.

These results are consistent with (but do not necessitate) a
picture in which the kinematics of the emission line gas are predominantly
dictated by gravity and the line luminosity of the nebula is predominantly
determined by the ionizing luminosity of the central engine.  If the mass
of cold gas increases with redshift,
then the increasingly powerful AGN present in the 3CR sample at higher
redshift, will illuminate a larger ionized nebula.
If the velocities are predominantly gravitational, then
the most powerful sources which are at the highest redshifts
in our sample, and show the highest
emission line gas velocity, would have to be
found in very deep gravitational potentials, such as cluster
environments. Alternatively, at the highest redshifts,
powerful radio galaxies may be in environments which are in a relatively young
evolutionary state where the kinematics are still dominated by infall.
In the gravitational interpretation, the very highest velocity systems,
with apparently disjoint multiple velocity systems in the nebula of a
given source, are cases where two galaxies are merging or interacting,
and gas systems associated with each galaxy are seen.

\section{Acknowledgments}

We thank the staffs of the Kitt Peak and Cerro Tololo sites of the National Optical Astronomy
Observatories as well as those of the Lick and Palomar Observatories. At the time these
observations were being carried out, the Carnegie Institution had access to the Hale
5m telescope under a cooperative agreement with the California Institute of Technology.


\begin{figure}
\epsfxsize=7in
\epsffile{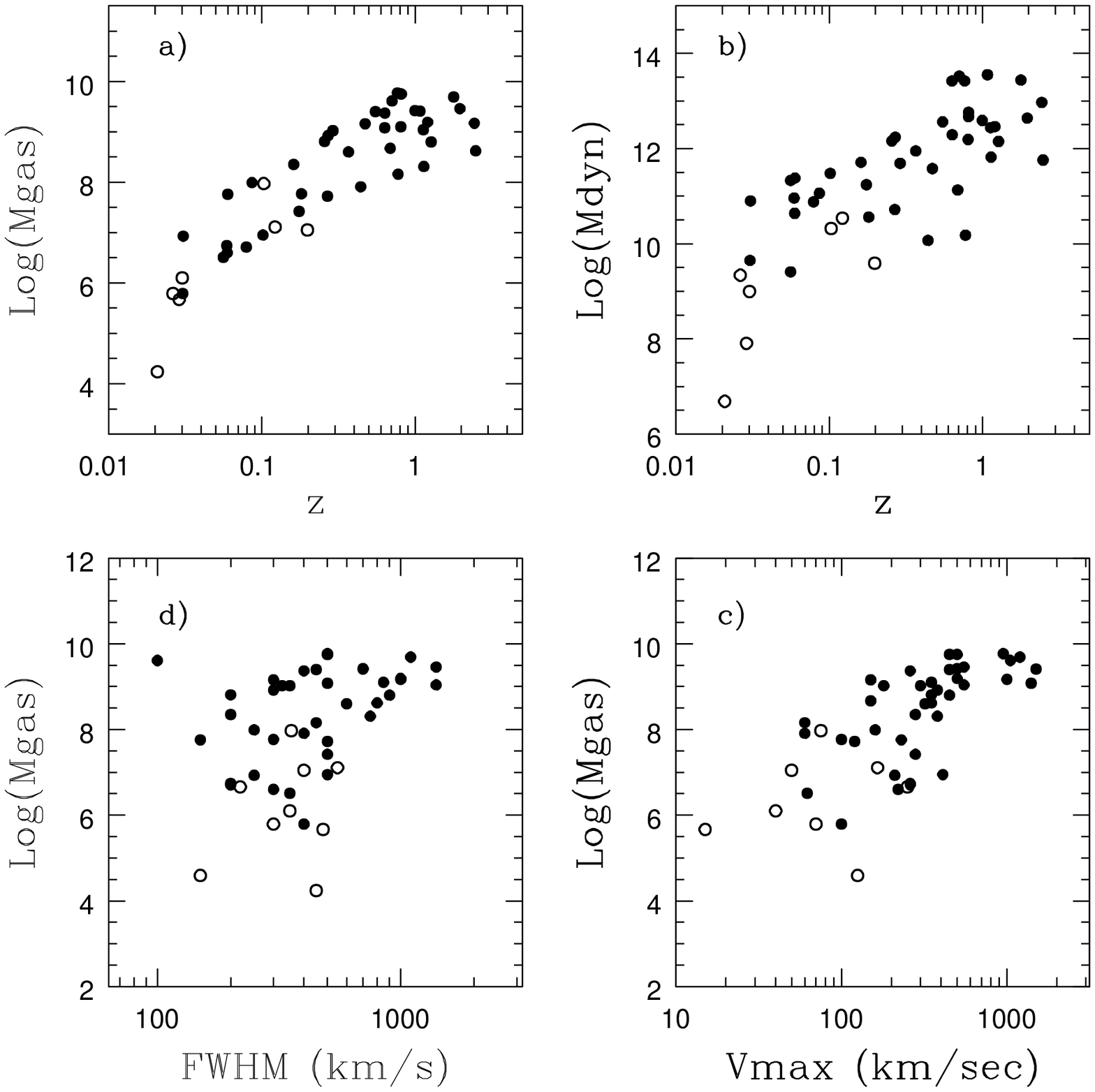}
\caption{In each of the panels of this figure, and the three subsequent
figures, the filled symbols refer to FRII sources, while the open symbols
represent FRI and transition sources. In the upper panels the mass of
ionized gas and the apparent dynamical mass are plotted against
redshift. In the lower panels the derived mass of ionized gas is plotted
against the maximum FWHM (left) and the maximum velocity (right). All of
the masses are shown in units of M$_{\odot}$.  } 
\end{figure}

\begin{figure}
\epsfxsize=7in
\epsffile{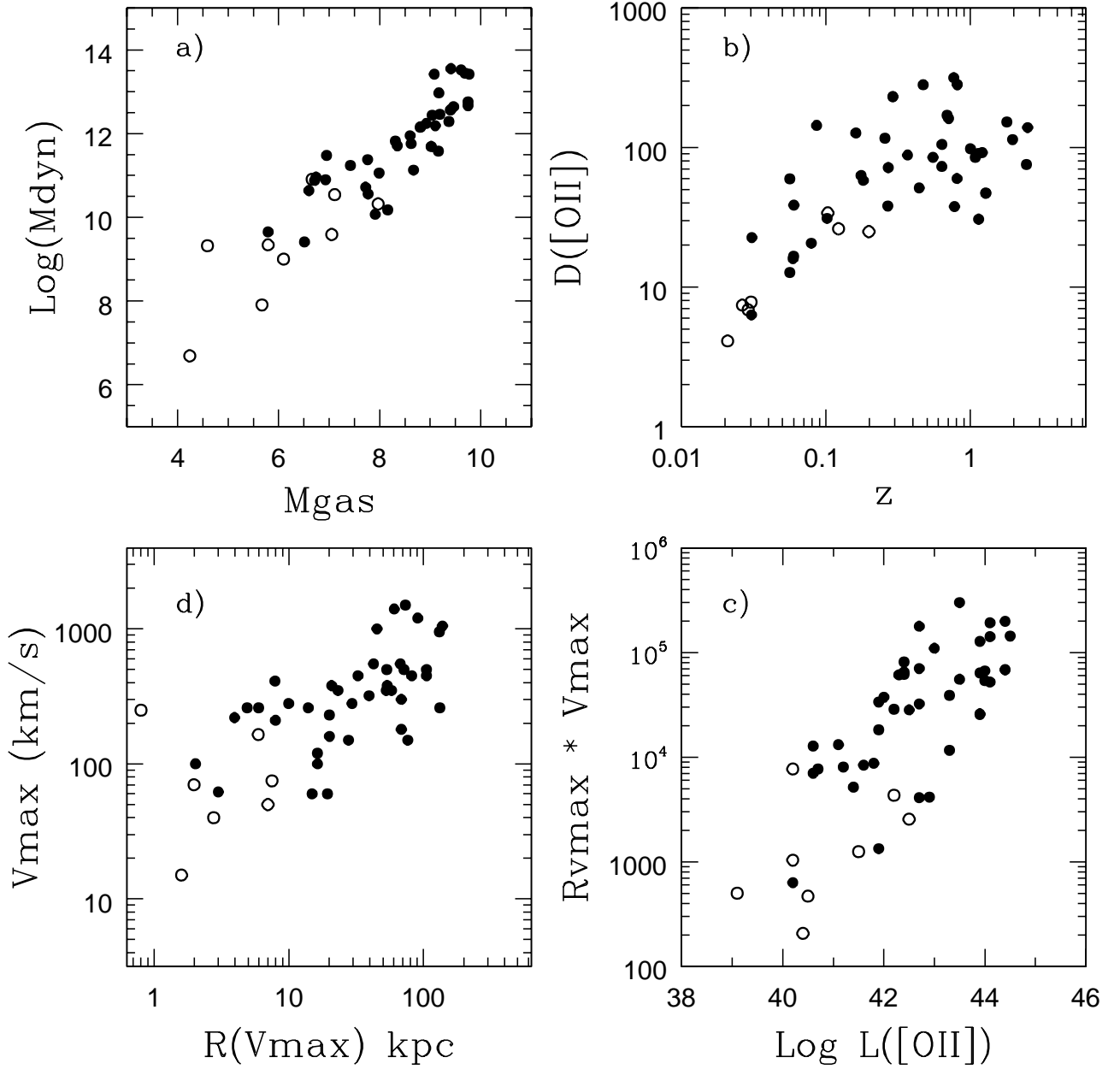}
\caption{In the upper left the apparent dynamical mass is plotted against the
derived mass of ionized gas. See the discussion in the text of the degeneracy between
these derived quantities. The isophotal emission-line size is plotted against redshift
in the upper right panel, while the maximum velocity is plotted against the radial
position of the fastest moving [OII] emission in the lower left. The lower right
panel shows the run of R$_{Vmax} \times V_{max}$ with emission-line luminosity.  } 
\end{figure}

\begin{figure}
\epsfxsize=7in
\epsffile{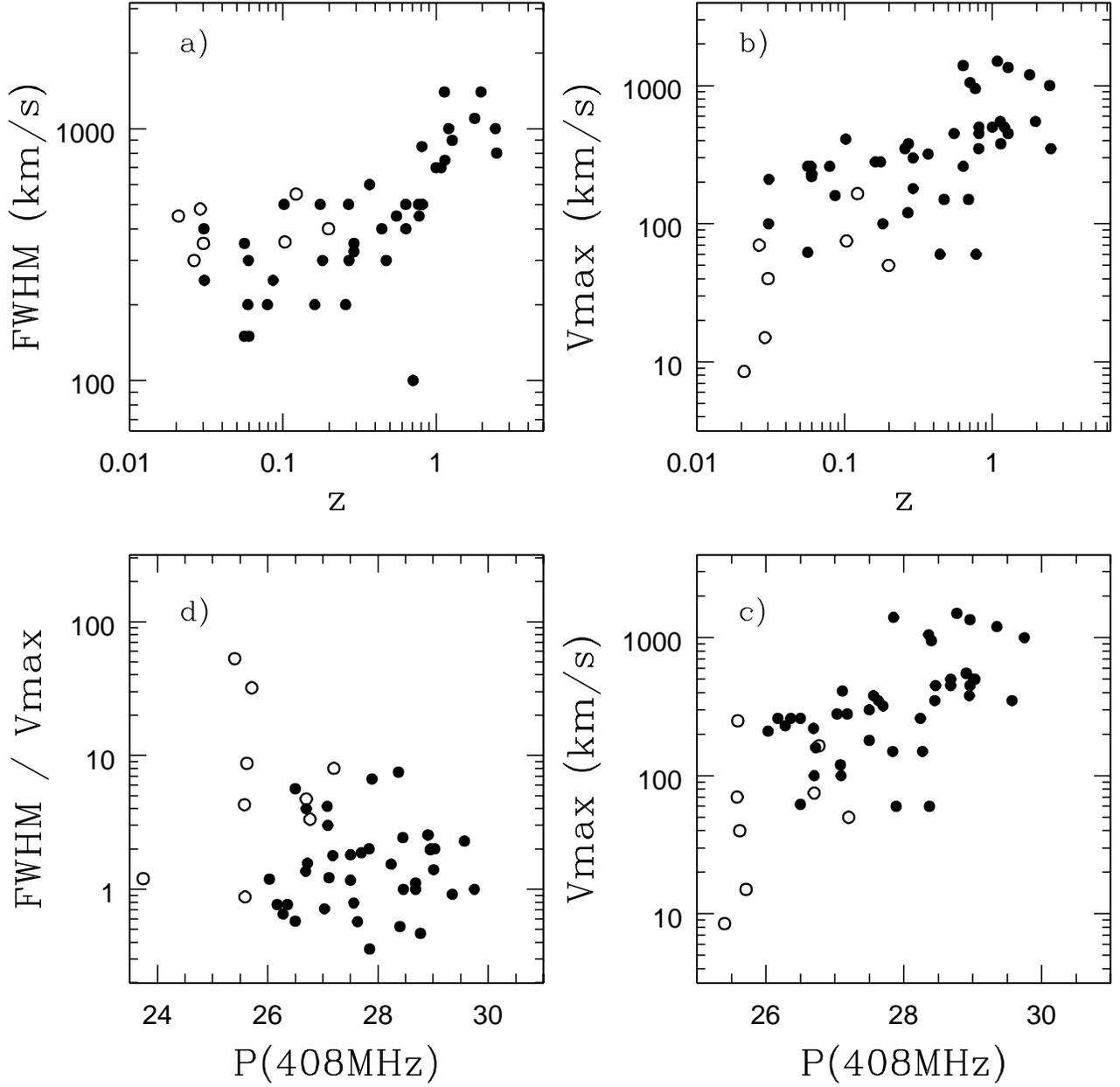}
\caption{The upper left and right panels show the run of
FWHM and V$_{Max}$ against redshift, while the lower panels
show the ratio of the FWHM to the maximum velocity (left) and
maximum (right) velocity against the 408MHz monochromatic radio power.
} 
\end{figure}

\begin{figure}
\epsfxsize=7in
\epsffile{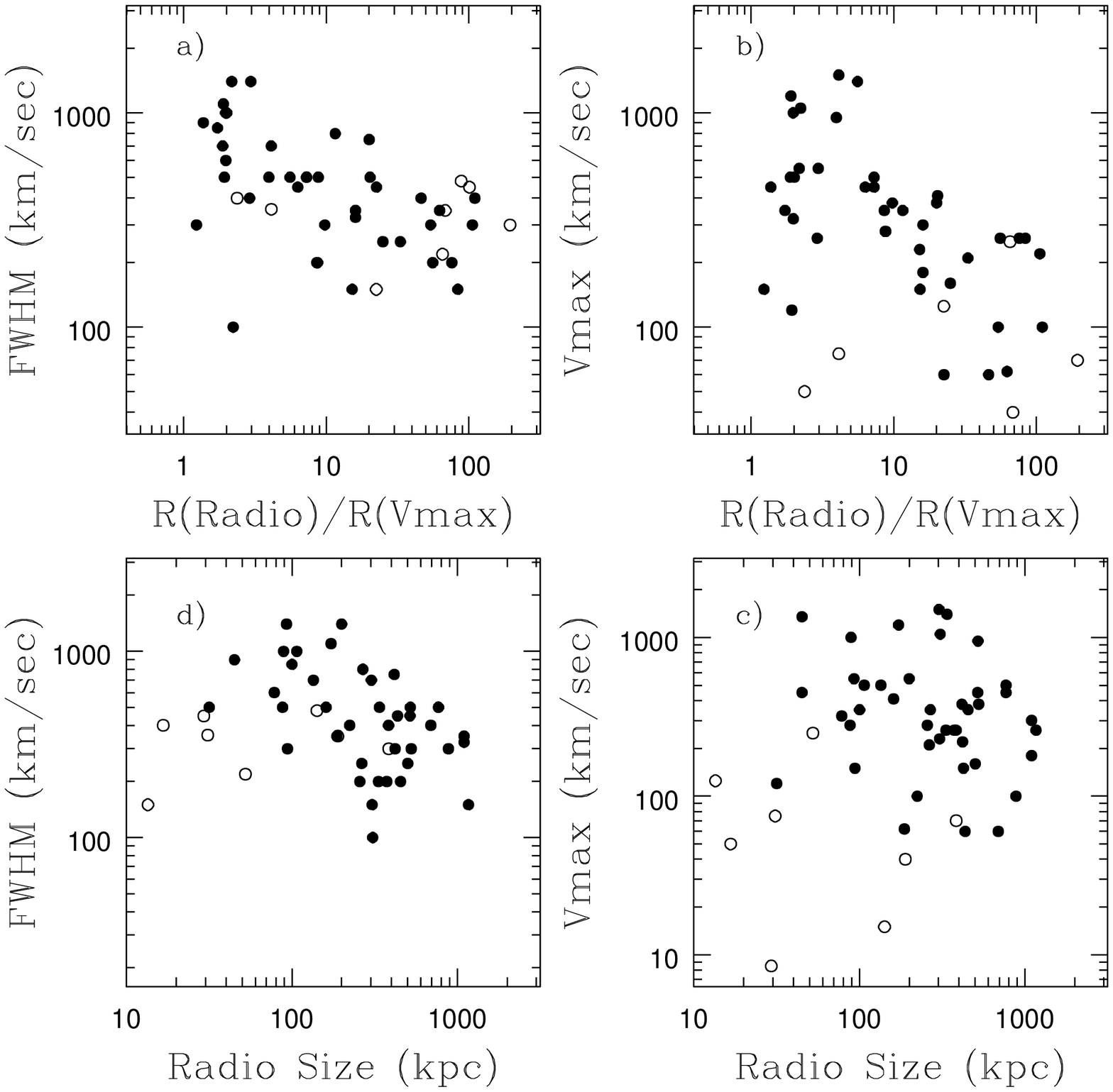}
\caption{Plotted are the off-nuclear FWHM (upper left) and the 
maximum velocity (upper right) against the
ratio of the radio and emission line sizes. The lower left
and right panels show the same quantities plotted against the
full linear size of the radio source.
} 
\end{figure}

\begin{thebibliography}

\bibitem{}{ Allen, S. W., Fabian, A. C., Kneib, J. P., 1996, MNRAS, 279, 616 }



\bibitem{}{Barthel, P. D., \& Miley, G. K., 1988, Nature, 333, 319}



\bibitem{}{ Baum, S. A., Heckman, T., Bridle, A., van Breugel, W.,
$\&$ Miley, G. 1988, ApJS,  68, 643 }

\bibitem{}{ Baum, S. A., $\&$ Heckman, T. 1989$a$,  \apj,  336, 681 }

\bibitem{}{ Baum, S. A., $\&$ Heckman, T. 1989$b$,  \apj,  336, 702. }

\bibitem{}{ Baum, S. A., Heckman, T., $\&$ van Breugel, W. 1992, ApJ,  389, 208 }

\bibitem{}{ Baum, S. A., Heckman, T., $\&$ van Breugel, W. 1990, ApJS,  74, 389 }

\bibitem{}{ Baum, S. A., Zirbel, E. L., \& O'Dea, C. P., 1995, ApJ, 451, 88}

\bibitem{}{ Best, P., Longair, M., \& Rottgering, H. 1999, in "The Hy-Redshift
Universe", ASP Conf. Series, A. Bunker \& W. van Breugel, eds. in press}


\bibitem{}{ Bennet, A. S. 1962, MNRAS, 68, 163 }

\bibitem{}{ Bicknell, G. \& Koekemoer, A. 1995, in IAU Symp 175,
``Extraglactic Radio Jets'', R. Ekers et al. eds., p 473.}

\bibitem{}{Bicknell, G., Dopita, M. A., \& O'Dea, C. P., 1997, ApJ, 485, 112}



\bibitem{}{ Capetti, A., Axon, D., Macchetto, F. D., Marconi, A.
Winge, C. 1999, ApJ, 516, 187 }


\bibitem{}{ Clark, N. E., Axon, D. J., Tadhunter, C. N., Robinson, A., O'Brien, P. 1998, ApJ, 494, 546 }




\bibitem{}{Djorgovski, S., Spinrad, H., McCarthy P., 
    Dickinson, M., van Breugel, W., $\&$ Strom, R. 1988, AJ, 96, 836 }



\bibitem{}{ Ellingson, E., Green, R., \& Yee, H. 1991, ApJ 371, 49}

\bibitem{}{ Fanaroff, B., Riley, J. M. 1974, MNRAS, 167, 31}





\bibitem{}{Gelderman, R. \& Whittle, M.  1994, ApJS, 91, 491}




\bibitem{}{ Heckman, T. M., Baum, S. A., van Breugel, W. J. M., \& McCarthy, P.
1989, ApJ, 338, 48}

\bibitem{}{ Hill, G. J., \& Lilly, S. J., 1991, ApJ, 367, 1}










\bibitem{} {Ledlow, M. J., \& Owen, F. N., 1996, AJ, 112, 9}






\bibitem{}{ Leahy, J. P., Bridle, A., \& Strom, R. G. 1995, 
in "Extragalactic Radio Sources", R. D. Ekers et al., eds., Kluwer, IAU Symp. 175, 157}





\bibitem{}{ Longair, M., et al. 1999, in "The Hy-Redshift Universe", ASP conference series,
in press.}


%


\bibitem{}{Mazure, A. \etal\ 1996, A\&A, 310, 31 }

\bibitem{}{McCarthy, P. 1993, Ann. Rev. A\&Ap, 31, 639 }


\bibitem{}{ McCarthy, P. J., van Breugel, W. J. M., 
Spinrad, H., \& Djorgovski, S. 1987,  \apj,  321, L29 }




\bibitem{}{McCarthy, P., van Breugel, W., \& Kapahi, V. 1991, ApJ, 371, 478 }

\bibitem{}{ McCarthy, P., Spinrad, H., \& van Breugel, W.,  1995, ApJS, 99, 27}

\bibitem{}{McCarthy, P. J., Baum, S. A., \& Spinrad, H. 1996a, ApJS, 106, 281}

\bibitem{}{McCarthy, P. J., Kapahi, V. K., van Breugel, W., Persson, S. E.,
Athreya, R., \& Subrahmanya, C. R., 1996b, ApJS, 107, 19}

\bibitem{}{Miley, G. K., Chambers, K. C., van Breugel, W., Macchetto, F. 1992, ApJ, 401, L69 }



\bibitem{}{Miralde-Escud\`e, J., \& Babul, A., 1995, ApJ, 449, 18}

\bibitem{}{O'Dea, C. P., 1998, PASP, 110, 493}








\bibitem{}{ Prestage, R. M., Peacock, J. A. 1988, MNRAS, 230, 131 }


\bibitem{}{ Rawlings, S. \& Saunders, R. 1991, Nature, 349, 138}

\bibitem{}{ Robinson, A., Binette, L., Fosbury, R. A. E., Tadhunter, C. N. 1987, MNRAS, 227, 97 }
 
 











\bibitem{}{Spinrad, H., Djorgovski, S., Marr, J., $\&$ Aguilar L. A. 1985, 
 PASP, 97, 932 }



\bibitem{}{Squires, G., Neumann, D. M., Kaiser, N., Arnaud, M., Babul, A., 
Boehringer, H., Fahlman, G., \& Woods, D., 1997, ApJ, 482, 648}


\bibitem{}{ Strom, R. G., Riley, J. M., Spinrad, H., van Breugel, W., Djorgovski, 
S., Liebert, J., $\&$ McCarthy, P.  1990, A\&A,  227, 19 }


\bibitem{}{Tadhunter, C. N., Fosbury, R. A. E., $\&$ Quinn, P. J., 1989, MNRAS,
240, 225}

\bibitem{}{ Tadhunter, C. N., Packham, C., Axon, D. J., Jackson, N. J., 
Hough, J. H., Robinson, A., Young, S., Sparks, W. 1999 MRNAS, 307, 24 }


\bibitem{}{Tadhunter, C. N., Morganti, R.,
Robinson, A., Dickson, R., Villar-Martin, M., \& Fosbury, R. A. E. 1999, 298, 1035}

\bibitem{}{Villar-Martin, M., Tadhunter, C. N., Clark, N.,
1997, A\&A, 323, 21}

\bibitem{}{Villar-Martin, M., Tadhunter, C. N., Morganti, R.,
Clark, N., Killeen, N., \& Axon, D. 1998, A\&A, 332, 479}

\bibitem{}{Villar-Martin, M., Binette, L., \& Fosbury, R. A. E. 1999, A\&A, 346, 7}

\bibitem{}{de Vries, {\it et al.} 1997, ApJS, 110, 191}

\bibitem{}{de Vries, W. H., O'Dea, C. P., Baum, S. A., Perlman, E., Lehnert, M. D.
\& Barthel, P. D. 1998, ApJ 503, 156}

\bibitem{}{Whittle, M., 1992, ApJ, 387, 109}

\bibitem{}{Xu, C., Livio, M. \& Baum, S., 1999, AJ, 118, 1169}



\bibitem{}{Yates, M., Miller, L., \& Peacock, J. 1989, MNRAS, 240, 129}

\end{thebibliography}
\end{document}